\documentclass[aps,final,notitlepage,oneside,twocolumn,nobibnotes,nofootinbib,
superscriptaddress,noshowpacs,centertags]{revtex4-1}

\usepackage[english]{babel}
\usepackage{graphicx}
\usepackage{latexsym}
\usepackage{amssymb}
\usepackage{amsmath}
\usepackage{float}

\usepackage{xcolor}

\begin{document}

\title{Production and evaporation of micro black holes as a link between mirror universes}
\author{Viktor K. Dubrovich}\thanks{e-mail: dvk47@mail.ru}
\affiliation{Special Astrophysical Observatory, St. Petersburg Branch, Russian Academy of Sciences, St.
Petersburg, 196140 Russia}
\author{Yury N. Eroshenko}\thanks{e-mail: eroshenko@inr.ac.ru}
\affiliation{Institute for Nuclear Research, Russian Academy of Sciences, pr. 60-letiya Oktyabrya 7a, Moscow, 117312 Russia}
\author{Maxim Yu. Khlopov}\thanks{e-mail: khlopov@apc.in2p3.fr}
\affiliation{Institute of Physics, Southern Federal University, Rostov on Don, Russia}
\affiliation{National Research Nuclear University ”MEPHI”
(Moscow Engineering Physics Institute), 115409 Moscow, Russia}
\affiliation{Centre for Cosmoparticle Physics “Cosmion” 115409 Moscow, Russia}
\affiliation{Universit\'e de Paris, CNRS, Astroparticule et Cosmologie, F-75013 Paris, France}

\date{\today}


\begin{abstract}
It is shown that the equalization of temperatures between our and mirror sectors occurs during one Hubble time due to microscopic black hole production and evaporation in particle collisions if the temperature of the Universe is near the multidimensional Plank mass. This effect excludes the multidimensional Planck masses smaller than the reheating temperature of the Universe ($\sim10^{13}$~GeV) in the mirror matter models, because the primordial nucleosynthesis theory requires that the temperature of the mirror world should be lower than ours. In particular, the birth of microscopic black holes in the LHC is impossible if the dark matter of our Universe is represented by baryons of mirror matter. It excludes some of the possible coexisting options in particle physics and cosmology. Multidimensional models with flat additional dimensions are already strongly constrained in maximum temperature due to the effect of Kaluza-Klein mode (KK-mode) overproduction. In these models, the reheating temperature should be significantly less than the multidimensional Planck mass, so our restrictions in this case are not paramount. The new constraints play a role in multidimensional models in which the spectrum of KK-modes does not lead to their overproduction in the early Universe, for example, in theories with hyperbolic additional space.
\end{abstract}

\maketitle



\section{Introduction}

The mirror matter model was proposed by Lee and Yang in \cite{LeeYan56} and developed in  \cite{KobOkuPom66} (see review and bibliography in \cite{Oku07,Khl11}). These models have many interesting consequences for cosmology and astrophysics \cite{BerCom01,Beretal05,BenBer01,Ber04}, in particular, the mirror dark matter can form objects of different types \cite{BliKhl82,BliKhl83,Khlea91}, including domain structures \cite{DubKhl89}.

In several works the possibility was considered that our and mirror worlds interact not only gravitationally, but also through some exchange of energy and matter. The matter can be transferred by the processes related to leptons \cite{BenBer01,Ber04,BenBer01-2,Ber05}, neutrons \cite{BerBen05,BerBen06,BerGaz11} or neutrinos \cite{AkhBerSen92,BerMoh95,BerVil00,BerNarVis03} with oscillations into the particles of mirror world and vice versa due to the high-order operators in the Lagrangian. The authors of \cite{CarGla87} considered the hypothesis about the mixing of our and mirror photons by the following term in Lagrangian $\varepsilon F_{\mu\nu}F'_{\mu\nu}$, and the restriction $\varepsilon<3\times10^{-8}$ was obtained from the primordial nucleosynthesis constraints. In more detail the evolution of the temperatures in our and in the mirror words due to the mixing of photons was considered in \cite{BerLep08,CiaFoo08,FooVag15}, where the constraints on the mixing parameter $\varepsilon$ were elaborated. Nonorientable wormholes provide another canal for the matter exchange \cite{DokEro14}.  

The formation of microscopic black holes (BHs) in particle collisions in the early Universe was discussed earlier in articles \cite{BarFerGra05,ConWiz07, BorMas10,NakYok18,SaiSto18,Baretal19}.  In this paper, we consider the energy exchange between our and the mirror world by the birth and evaporation of microscopic BHs. As far as we know, previously this energy exchange channel with reference to mirror matter was not considered. The connection of the worlds through microscopic BHs was considered in other aspects in the works \cite {DvaPuj09}, \cite {DvaMic09}, where it is shown that microscopic black holes can provide bridges between close branes.

Primordial nucleosynthesis requires that the temperature of the CMB in the mirror world be lower than in ours \cite{BerDolMoh96}. Otherwise, additional relativistic degrees of freedom appear, which change the dynamics of the primordial nucleosynthesis and change the yield of chemical elements. Can the temperatures of our and the mirror worlds in the early Universe be leveled by the exchange of energy between them? Aforementioned variant with the photons mixing have already been considered. In the presence of additional dimensions,  multidimensional Planck mass $M$ can be many orders of magnitude smaller than the usual 4-dimensional Planck mass, which reduces the energy necessary for the BH production. This effect was widely discussed in relation to the Large Hadron Collider. If a BH is born when two particles of our world collide, the BH evaporates both in our and in mirror particles. Thus, there is a transfer of energy from our to the mirror world. The reverse flow of energy will be less, because the temperature of the mirror world is lower. Because of the energy exchange, the temperatures of our and the mirror world can be equalized. This contradicts the primordial nucleosynthesis constraint and implies a lower bound on the Planck mass in multidimensional models. The main result of this work is the restriction on the multidimensional Planck mass. Namely, it was obtained that the multidimensional Planck mass should exceed the reheating temperature $\sim10^{13}$~GeV .

With the temperature decrease, the Universe evolves from multidimensional to our 4D state. The transition takes place near $T\sim  M$. We will show that the temperature equalization between our and mirror worlds occurs during one Hubble time. This means that to state the fact of the temperature equalization one has no need to consider the temperatures at $T>M$ and the complicated multidimensional physics. It's enough to consider only the $T\sim  M$ epoch. In this epoch the usual 4-dimensional physics are at place. Therefore, our consideration will be in much respects independent on the particular models of the multidimensional word at $T>M$.

Note that, in the models with flat extra-dimensions, the reheating temperature is significantly lower than the multidimensional Planck mass due to cosmological constraints (KK-mode overproduction) \cite{Han01}. In this regard, our calculations are not applicable to all the extra-dimension models, but only to those where there are no cosmological constraints (see Discussion in \cite{Han01}). These are models with a hyperbolic compact manifold \cite{Kal00} and models with many branes. In these models the cosmological bounds disappear completely.


\section{Preliminary estimates}

Consider the cosmological model where our dark matter is a mirror substance (mirror baryon, mirror leptons etc), and the temperature of our world is different (higher) from the temperature of the mirror world. At temperatures $T\geq M$, BHs will be born in the particle collisions, and in a world with a higher temperature, their birth is more efficient. During the quantum evaporation and decay of the BH, particles of both our and the mirror Universe are equally likely to be born. Thus, there will be a flow of energy from our hotter Universe to the colder mirror universe. Below, we will evaluate how effective this process is, and whether the temperatures of our world and the mirror world will equalize.

With the additional dimensions present, the multidimensional Planck mass $M$ can be less than the usual 4-dimensional Planck mass $M_{\rm Pl}\simeq1.2\times10^{19}$~GeV. If $M\sim10$~TeV the creation of microscopic BHs at the Large Hadron Collider  is possible \cite{DimLan01,Kin02,CMS11}. 
The birth of microscopic BHs in two particles collision can occur under the following two conditions. (i) The energy of the particles in the center of mass system is of the order or greater than the multidimensional Planck mass $M$. (ii) Colliding particles should approach one another to a distance less than the multidimensional Schwarzschild radius. In this case, the energy carried by the particles will be enclosed under the gravitational radius, and the formation of a black hole happens. The cross section for the production of BHs in $pp$ collisions in the first approximation is written in the form \cite{BanFis99,DimLan01}, 
\begin{equation}
\sigma\simeq\pi R_s^2=\frac{1}{M^2}\left[\frac{M_{\rm BH}}{M}\left(\frac{8\Gamma\left(\frac{n+3}{2}\right)}{n+2}\right)\right]^{2/(n+1)},
\label{sig1}
\end{equation}
where $n$ is the number of additional dimensions ($4+n$ in total), $R_s$ is the Schwarzschild radius of the multidimensional BH with mass $M_{\rm BH}$, and the units $\hbar=c=1$ are in use here and further. The Boltzmann constant is also assumed to be equal 1.

Eq.~(\ref{sig1}) gives the cross section in the case of the flat additional dimensions. The solution similar to the Schwarzschild solution, in the case of a hyperbolic additional space may differ on scales larger than the curvature scale of the additional space. For the purposes of this article, however, it is sufficient to consider situations where the Schwarzschild radius is less or of the order of the curvature radius. Indeed, as it was shown in \cite{Kal00}, the minimum possible curvature radius is $\sim 1/M$. The Schwarzschild radius is $\leq 1/M$, since we consider only the stage of the Universe evolution when black holes with $M_{BH}\leq M$ are born (from the Eq.~(\ref{sig1}) it can be seen that then $R_s\leq 1/M$). Thus, we can use Eq.~(\ref{sig1}) as an estimate for the production cross section.

In different theories, the multidimensional (fundamental) Planck mass $M$ and 4-dimensional Planck mass $M_{\rm Pl}$ are connected in different ways through the volume of additional space. However, this mass relation will not be used in our calculations. We leave the $M$ as a free parameter without a specific type of connection with $M_{\rm Pl}$. For this reason, our calculations are practically independent of the topology of the additional space, and our results are applicable by order of magnitude. The only condition is the absence of cosmological restrictions associated with the birth of KK-modes. 

Let us begin from the simple estimates. Assume that for $T\sim M$ all BHs are born with masses $M_ {\rm BH}\sim M$. In reality, the mass spectrum should be formed. Assume also that the BHs decay immediately after birth without a stable remnants (or Plankions). In reality, the Hawking evaporation takes some time, and there is a time delay from the moment of birth to the moment of final decay. Taking into account the 1st assumption, the birth cross-section (\ref{sig1}) is written as
\begin{equation}
\sigma\sim\frac{2}{M^2}.
\label{sig2}
\end{equation}
Let one and only one BH appear with the cross-section (\ref{sig2}) when any two particles collide at $T\geq M$, and this takes place for each effective degree of freedom (total of $g_*\sim100$).

The multidimensional Planck mass $M$ appears only in the BH birth cross-section, and the usual 4-dimensional Planck mass $M_{\rm Pl}$ is used for the universe evolution.
The total number density of particles in the universe at the radiation-dominated stage \cite{GorRub08}
\begin{equation}
n\simeq g_*\frac{\rho}{3T},
\end{equation}
where
\begin{equation}
\rho=\frac{\pi^2}{30}g_*T^4.
\end{equation}
The number of BHs born in a volume $V$ per unit time is 
\begin{equation}
\dot N\sim \sigma n^2V.
\end{equation}
The products of BH quantum evaporation immediately thermalize, going into the cosmic plasma. The relative rate of energy transfer to the mirror universe 
\begin{equation}
\frac{\dot\rho_{\rm BH}}{\rho}\sim\frac{1}{\rho}\frac{M\dot N}{2V},
\label{dorrhorho}
\end{equation}
where the multiplier $1/2$ is associated with the equal probability of our and mirror particles birth during the BH decay. 

Compare (\ref{dorrhorho})
with the rate of the Universe expansion 
\begin{equation}
H=\frac{\dot a}{a}=\frac{T^2}{M_{\rm Pl}^*},
\end{equation}
where \cite{GorRub08} $M_{\rm Pl}^*=M_{\rm Pl}/(1.66\sqrt{g_*})$,
\begin{equation}
\left(\dot\rho_{\rm BH}/\rho\right)/H\sim2\times10^3\left(\frac{M_{\rm Pl}}{M}\right)\gg1.
\label{dotrhoest}
\end{equation}
Thus, the multidimensional Planck mass $M$ must be larger than the reheating temperature $\sim10^{13}$~GeV. Otherwise, the temperatures of our world and the mirror world equalize, and the Universe becomes symmetrical, violating the primordial nucleosynthesis constraints.


\section{The rate of micro black hole production in the early universe}

Let us consider the micro black hole production rate more exactly. The rate of particle interactions (number of events per time interval $dt$ inside the volume element $dV$) is expressed through the invariant cross-section $\sigma$ \cite{LL-2}
\begin{equation}
\frac{d\nu}{dtdV}=\sigma\frac{\sqrt{(p_1^\mu p_{2\mu})^2-m_1^2m_2^2}}{E_1E_2}n_1n_2,
\label{nueq}
\end{equation}
where $n_1$ and $n_2$ are the number densities of the colliding particles. The generalization of this equation for the physical situation in the early Universe requires the integration over particle distributions and the heat production looks as 
\begin{eqnarray}
\frac{\delta Q}{dtdV}&=&\frac{1}{2}\sum\limits_{i,j}\int d^3p_1\int d^3p_2\sigma\Delta E\frac{\sqrt{(p_1^\mu p_{2\mu})^2-m_1^2m_2^2}}{E_1E_2}
\nonumber
\\
&\times&\frac{1}{(2\pi)^6}\frac{1}{\left[e^{(E_1-\mu_1)/T}+i\right]\left[e^{(E_2-\mu_2)}/T+j\right]},
\label{doubleint}
\end{eqnarray}
where $\Delta E$ is the energy transferred into BH, and indexes $i,j$ are equal to $1$ in the case of fermion particles, and $-1$ for bosons. One has four possible combinations: $i=j=1$, $i=j=-1$, $i=-j=1$, and $i=-j=-1$.
The summation goes over all possible degrees of freedom (bosonic and fermionic). The factor $1/2$ takes into account the double count in the sum due to particle exchange. 

As will be shown later, the temperature equalization of our and the mirror universe is possible in just one Hubble time near $T\sim M$. In this sense, using the exact expressions for Fermi-Dirac and Bose-Einstein distributions seems redundant. However, it may be justified by the following reasons. At $T\ll M$ the microscopic BHs are not born in the collisions of most particles. But the distributions continue toward higher energies, and rare particles with energies $E\sim M$ from these distributions can produce microscopic BH in collisions. It cannot be excluded in advance that even a small fraction of all particles with energies $E\sim M$ will lead to the temperature equalization. Therefore, we keep the distributions in our calculations.

For the approximate calculation of (\ref{doubleint}) we do the following simplifications. First of all, we consider sufficiently high temperatures $T$ and ultrarelativistic case by neglecting $m_1$ and $m_2$ in the further equations. In this case $E_1=|\vec p_1|$, $E_2=|\vec p_2|$ and
\begin{equation}
p_1^\mu p_{2\mu}=|\vec p_1||\vec p_2|(1-\cos\theta),
\end{equation}
where $\theta$ is the angle between $\vec p_1$ and $\vec p_2$. 
Let us introduce the dimensionless variables
\begin{equation}
u_1=\frac{|\vec p_1|}{T}, \quad u_2=\frac{|\vec p_2|}{T},
\end{equation}
then
\begin{equation}
\Delta E=T(u_1+u_2).
\end{equation}
The chemical potential $\mu_1=\mu_2=0$ because of fast thermalization in hot plasma. Really, the thermalization of the evaporated radiation proceeds very fast. Sunyaev and Zeldovich have shown in \cite{SunZel70} that if the energy injection takes place prior to the epoch of the $e^+e^-$ pairs annihilation no observable distortions are expected in the spectrum of primordial radiation. It was obtained in \cite{IllSiu75} that even significant energy release at the red-shifts $z\geq10^8$ would be completely thermalized. Therefore, we use the thermal distributions for bosons and fermions.
We consider only the process $\chi_1+\chi_2\to $~BH neglecting the possible additional canals of the type $\chi_1+\chi_2\to $~BH~$+$ something else. In particular, we suppose, that gravitational waves generated during the particles collisions carry out energy of the order or less than $\Delta E$. Under this condition our calculations are valid at the order of magnitude at least. The center of mass energy squared is
\begin{equation}
s=(p_1+p_2)^2=2T^2 u_1u_2(1-\cos\theta)=M_{\rm BH}^2\geq\gamma^2M^2,
\end{equation}
where the factor $\gamma$ in the last inequality follows from the entropy arguments, and $\gamma\sim5$ \cite{Kin02}. The following condition is necessary for the above inequality to be satisfied
\begin{equation}
u_1u_2\geq\xi_{\rm min}=\frac{\gamma^2M^2}{2T^2}.
\end{equation}

\begin{figure}
	\begin{center}
\includegraphics[angle=0,width=0.45\textwidth]{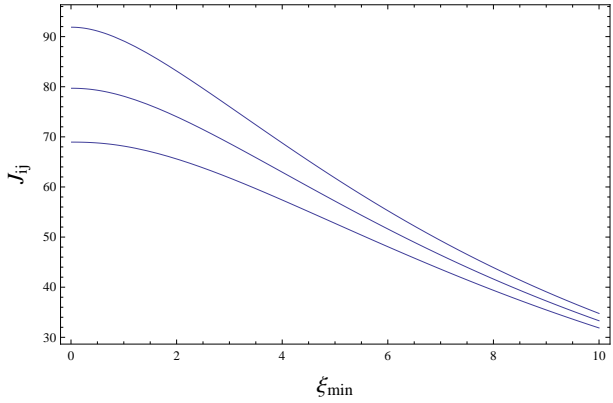}
	\end{center}
\caption{The functions $J_{ij}$ at $n=1$ in the cases (from up to down) $i=j=-1$, $i=-j=1$, and $i=j=1$.}
	\label{figjij}
\end{figure}

The invariant cross-section $\sigma$ is calculated in the laboratory system where one of the particle is at rest. Note however that during the transition to the center of mass system the $\sigma$ does not change and is given by (\ref{sig1}) because this geometrical cross-section represents the transverse direction under the Lorentz transformations. Note in addition that $dtdV$ in (\ref{nueq}) is invariant. For the massless particles the aforementioned laboratory system should be considered in the limit $m\to0$. We don't consider the possible exponential suppression of the geometrical cross-section which was proposed in \cite{Vol01}  and initiated discussion in several works.

The integration over the angle $\theta$ in (\ref{doubleint}) can be done analytically. After this the (\ref{doubleint}) takes the form
\begin{equation}
\frac{\delta Q}{dtdV}=\Phi T^{(7n+9)/(n+1)}
\label{dqdtdvfirst}
\end{equation}
where 
\begin{equation}
\Phi=J\frac{n+1}{2n+3}\frac{2^{1/(n+1)}}{(2\pi)^4M^{(2n+4)/(n+1)}}\left(\frac{8\Gamma\left(\frac{n+3}{2}\right)}{n+2}\right)^{2/(n+1)},
\label{phivbig}
\end{equation}
\begin{equation}
J=\sum J_{ij},
\end{equation}
\begin{eqnarray}
J_{ij}&=&\int\limits_0^\infty du_1\int\limits_0^\infty du_2(u_1+u_2)\left\{(u_1u_2)^{(2n+3)/(n+1)}\right.
\nonumber
\\
&-&\left.\xi_{\rm min}^{(2n+3)/(n+1)}\right\}\frac{\theta_H(u_1u_2-\xi_{\rm min})}{(e^{u_1}+i)(e^{u_2}+j)},
\end{eqnarray}
where $\theta_H$ is the Heaviside step function. 
The examples of these functions are shown at Fig.~\ref{figjij}. It's easy to see that $J_{ij}\to const$ at $\xi_{\rm min}\to0$, i.e. at $T\gg M$.


\section{Temperature evolution}

   \subsection{General equations}

Let us consider the evolution in time of the densities (or temperatures) of radiation in our and mirror universes. The values related to our world are marked by lower index of ``1'', and the values related to the mirror world are marked by ``2''. The first necessary equation is one of the Friedmann equations
\begin{equation}
\frac{1}{2}\left(\frac{da}{dt}\right)^2-\frac{4\pi G}{3}a^2(\varepsilon_1+\varepsilon_2)=-\frac{k}{2},
\end{equation}
where the densities enter in the form of a simple sum according to the summation of energy-momentum tensors in Einstein's equations, and further we consider a flat model with $k=0$.
The energy densities of our and mirror worlds are expressed through their temperatures
\begin{equation}
\varepsilon_1=g_*(T_1)\frac{\pi^2}{30\hbar^3c^3}T_1^4,\qquad \varepsilon_2=g_*(T_2)\frac{\pi^2}{30\hbar^3c^3}T_2^4,
\end{equation}
where $g_*(T)$ is the effective number of degrees of freedom.

Note that the multidimensional Planck mass $M$ enters only the BH production cross section, and the usual four-dimensional Planck mass $M_{\rm Pl}$ is used in the cosmological evolution equations, because the Einstein equations contain the already reduced gravitational constant at $T\ll M$.

Let us write down the first law of thermodynamics for the matter of our world:
\begin{equation}
\delta Q_1=p_1dV+dE_1,
\end{equation}
where $\delta Q_1$ is the energy change in the volume $V$ due to the energy transfer to the mirror world and due to the reverse energy flow, $E_1=\varepsilon_1 V$, pressure $p_1=\varepsilon_1/3$.
We write the cubic volume element in the form $V=a^3r^3$. For a fixed comoving volume ($r=const$) one has
\begin{equation}
\frac{\delta Q_1}{Vdt}=3\frac{\dot a}{a}(p_1+\varepsilon_1)+\dot\varepsilon_1=4\frac{\dot a}{a}\varepsilon_1+\dot\varepsilon_1.
\label{term1}
\end{equation}
And the similar relationship holds for the mirror world
\begin{equation}
\frac{\delta Q_2}{Vdt}=3\frac{\dot a}{a}(p_2+\varepsilon_2)+\dot\varepsilon_2=4\frac{\dot a}{a}\varepsilon_2+\dot\varepsilon_2,
\label{term2}
\end{equation}
with $\delta Q_2=-\delta Q_1$. Here we neglect the energy that is stored in black holes at any time prior to evaporation, assuming that the evaporation takes place very quickly.

Summing (\ref{term1}) and (\ref{term2}), we obtain for the total values $\varepsilon=\varepsilon_1+\varepsilon_2$ and $p=p_1+p_2$ the usual relation
\begin{equation}
\frac{d\varepsilon}{d t}=-3\frac{\dot a}{a}(p+\varepsilon)
\label{sum}
\end{equation}
with known solutions \cite{GorRub08} 
\begin{equation}
\varepsilon=\frac{3c^2}{32\pi Gt^2}, \qquad a(t)\propto t^{1/2}.
\label{332piGt2}
\end{equation}

In the general case, the effective number of degrees of freedom $g_*$ depends on the temperature. However, we consider temperatures $T\geq1$~ TeV. At such temperatures, in the Standard Model of elementary particles and in the minimal supersymmetric model, one can assume $g_*=const$ because the new degrees of freedom are not excited with the temperature increase. Limiting values at high temperatures are $g_*=106.75$ and $g_*=228.75$, respectively, in the Standard Model and in MSSM \cite{Sch03}. In any case, we assume that  $g_*=const$ in the finite temperature range. This is true if the temperature equalization occurs fairly quickly at times on the order of the Hubble time, so that $g_*=const$ is a good approximation. 
Let us denote $\theta_1=T_1^4$ and $\theta_2=T_2^4$. In this case, the equations (\ref{term1}) and (\ref{term2}) take the form
\begin{equation}
\dot\theta_1+4\frac{\dot a}{a}\theta_1=\frac{\Phi}{2\alpha}(\theta_2^{(7n+9)/4(n+1)}-\theta_1^{(7n+9)/4(n+1)}).
\label{term1theta}
\end{equation}
\begin{equation}
\dot\theta_2+4\frac{\dot a}{a}\theta_2=\frac{\Phi}{2\alpha}(\theta_1^{(7n+9)/4(n+1)}-\theta_2^{(7n+9)/4(n+1)}),
\label{term2theta}
\end{equation} 
where $\alpha=g_*\pi^2/(30\hbar^3c^3)$. 

   \subsection{Finite lifetime}

The important question is the finite lifetime of BH in the Hawking evaporation. Above we considered the instantaneous decay of BHs. Now we discuss the influence of the time-delay. The BHs lifetime is estimated as \cite{Kin02}
\begin{equation}
\tau\sim\frac{\hbar}{Mc^2}\left(\frac{M_{\rm BH}}{M}\right)^{(n+3)/(n+1)}.
\label{lifetime}
\end{equation} 
For the moving BH the additional Lorentz-factor $\Gamma\sim E/M_{\rm BH}$ arises in the lifetime (\ref{lifetime}). But typically $E\sim T$, $M_{\rm BH}\sim T$, and $\Gamma\sim 1$.
Let us compare this lifetime with cosmological (Hubble) time in the early Universe in the case $n=1$
\begin{equation}
\frac{\tau}{t}\sim2.7\times10^{-5}\left(\frac{T}{M}\right)^3\left(\frac{T}{10^{13}\mbox{~GeV}}\right)^{-1}.
\end{equation}
Therefore, the condition $\tau<t$ requires
\begin{equation}
M>3\times10^{-2}T\left(\frac{T}{10^{13}\mbox{~GeV}}\right)^{-1/3},
\label{tau5}
\end{equation}  
and in this case the rough condition for the BHs production $T\geq M$ can be satisfied only in the temperature range
\begin{equation}
T=(2.7\times10^{-5}-1)\times10^{13}\mbox{~GeV}.
\label{tau6}
\end{equation}  
Under the conditions (\ref{tau5}) and  (\ref{tau6}) the BHs decay typically during one Hubble time and the energy transfer between our and mirror universes can be considered as instantaneous. In this case we can use the expression (\ref{dqdtdvfirst}) for the energy transfer.  For $n>1$ the lifetime (\ref{lifetime}) becomes even shorter and the above conditions become softer.  

Otherwise one should use the integro-differential equations for the description of the energy transfer with time-delay. In this paper we do non use such approach for the following simple reason. We want to derive some lower bound on the $M$. The time-delay makes the energy transfer even more effective, because the radiation energy of the evaporated BH is red-shifted and diluted as $1/a^4(t)$ during the universe expansion. But the energy stored in the non-relativistic part of the BHs spectrum is rarefied  slowly as $1/a^3(t)$. Therefore, neglecting the finite lifetime we will obtain the lower limit for energy transfer which is enough for our purposes. 

We assume also that the BHs evaporate without stable remnants (Planckions).

   \subsection{Case $n=1$}

In the $n=1$ case the exact analytical solution can be found. Note however that the case $n=1$ is excluded by Newtonian law at Solar System distances
\cite{ArkDimDva98}. The equations (\ref{term1theta}) and (\ref{term2theta}) have the form
\begin{equation}
\dot\theta_1+4\frac{\dot a}{a}\theta_1=\frac{\Phi}{2\alpha}(\theta_2^2-\theta_1^2).
\end{equation}
\begin{equation}
\dot\theta_2+4\frac{\dot a}{a}\theta_2=\frac{\Phi}{2\alpha}(\theta_1^2-\theta_2^2).
\end{equation}
Now we take the difference of these equations. The right-hand-side can be decomposed $\theta_2^2-\theta_1^2=(\theta_2-\theta_1)(\theta_2+\theta_1)$ and the general expression (\ref{332piGt2}) can be used for the sum $\theta_2+\theta_1$. The resultant equation for $\theta_1-\theta_2$ have the simple exact solution. Let us also denote 
\begin{equation}
\delta=\frac{\theta_1-\theta_2}{\theta_1+\theta_2}.
\end{equation}
At the time of reheating $\delta=\delta_i\leq1$, and the maximum $\delta_i=1$ corresponds to a completely  cold or empty mirror universe.
With the initial condition $\delta(t_i)=\delta_i$ one has the solution
\begin{equation}
\delta(t)=\delta_i\exp\left\{\frac{3c^2\Phi}{32\pi G\alpha^2}\left(\frac{1}{t}-\frac{1}{t_i}\right)\right\}.
\end{equation}
We require that at $\delta_i\sim1$ and $t\gg t_i$ the situation $\delta(t)\ll1$ does not occur. One has numerically
\begin{equation}
\frac{3c^2\Phi}{32\pi G\alpha^2}\frac{1}{t_i}\simeq90\left(\frac{T_i}{10^{13}\mbox{~GeV}}\right)^{2}\left(\frac{M}{10^{13}\mbox{~GeV}}\right)^{-3}.
\label{estn1}
\end{equation}
Let us consider the temperatures $T_i\sim M$. We see that (\ref{estn1}) is less then 1 for $M>9\times10^{14}$~GeV. Otherwise the temperatures equalization take place during one Hubble time. In other words, the mass $M$ cannot be less than the reheating temperature as long as the reheating temperature of the Universe is less than $\sim10^{15}$~GeV.

   \subsection{General case}

For $n\geq2$, the exact solution of the equations (\ref{term1theta}) and (\ref{term2theta}) cannot be found, but, nevertheless, one can obtain a sufficiently strong lower bound on $M$. Taking again the difference (\ref{term1theta}) and (\ref{term2theta}), we obtain the equation
\begin{equation}
\frac{d}{dt}(\theta_1-\theta_2)+\frac{2}{t}(\theta_1-\theta_2)=-\frac{\Phi}{\alpha}(\theta_1^{(7n+9)/4(n+1)}-\theta_2^{(7n+9)/4(n+1)}).
\label{diftheta}
\end{equation}
If we replace the right-hand side of (\ref{diftheta}) by a smaller quantity (by absolute value), then the resulting equation will describe the process of energy transfer with lower efficiency than the original equation, and from the properties of its solution we obtain a lower bound on $M$.
We will consider the equation (\ref{diftheta}) in the bounded temperature range $T_i>T_2>T_f$, where $T_f$ will be chosen later. Note, that
\begin{equation}
\beta=\frac{7n+9}{4(n+1)}=1+\frac{3}{4}+\frac{1}{2(n+1)}.
\end{equation}
Note also that the function $\phi(x)=x^\beta$ at  $\beta>1$ is convex downward and for this case one can write
\begin{equation}
\phi(x_1)-\phi(x_2)>(x_1-x_2)\phi'(x_2)=(x_1-x_2)\beta x_2^{\beta-1}
\label{vyp}
\end{equation}
at $x_1>x_2$. As $x$, we take $x_{1,2}=T_{1,2}/M>1$. In the case $x_2>1$ the right-hand side of (\ref{vyp}) decreases even more 
if $1/[2(n+1)]$ is thrown out the exponent $\beta-1$. Therefore, we replace the equation (\ref{diftheta}) by the following
\begin{equation}
\frac{d}{dt}(\theta_1-\theta_2)+\frac{2}{t}(\theta_1-\theta_2)=-\frac{7\tilde\Phi M}{4\alpha}(\theta_1-\theta_2)\left(\frac{T_f}{M}\right)^3,
\label{difthetaappr}
\end{equation}
where $\tilde\Phi=4.6\times10^{-3}J$ is obtained after the minimization of (\ref{phivbig}).
Solving (\ref{difthetaappr}), we find for the relative change
\begin{equation}
\delta(t)=\delta_i\exp\left\{-\frac{7\tilde\Phi M}{4\alpha}\left(\frac{T_f}{M}\right)^3(t-t_i)\right\}.
\label{solgen}
\end{equation}

With the effective temperature equalization near $T_2\sim T_f$, we have the situation $T_1\sim T_2$. Let us consider the temperature variation during one Hubble time after $t_i$, i.e. we set again $M\sim T_f\sim T_i$. Also by order of magnitude $J\sim80g_*^2$. Under these conditions, the exponent in (\ref{solgen}) at $t=t_f\sim2t_i$ is 
\begin{equation}
\frac{7\tilde\Phi M t_i}{4\alpha}\simeq7\times10^5\left(\frac{M}{10^{13}\mbox{~GeV}}\right)^{-1}.
\label{itoggen}
\end{equation}
We see that in this case the temperature equalisation occurs at all masses $M$ during one Hubble time. The only way to avoid it is to suppose that $M$ is larger than the maximum temperature in the history of the hot Universe, i.e. the reheating temperature. Therefore the effect of the microscopic BH production excludes the masses $M<T_r\sim10^{13}$~GeV in the mirror matter models.


\section{Conclusion}

In this paper the implications of micro black hole formation in high-energy particles collisions for the mirror matter cosmologies are considered. Multidimensional Planck mass $M$ can be less than the usual 4-dimensional Planck mass easing the micro black holes production. Consider the model of the universe with two sectors: our usual sector and the mirror one. The temperature of our world should be higher then the temperature of the mirror one due to the primordial nucleosynthesis constraints. The production of the microscopic black holes is more efficient in a world with a higher temperature. During the quantum evaporation of  black holes, particles of both our and mirror universes will be emitted with equal probability. Thus, there will be a flow of energy from our  hotter Universe to the colder mirror world, and the equalization of their temperatures is possible. This effect allows to obtain the constraints on the multidimensional Planck mass $M$ in the mirror matter model. Namely, $M$ should be larger than the reheating temperature $\sim10^{13}$~GeV -- the maximum temperature in the hot Universe. Otherwise, the temperatures would be equalized and the  primordial nucleosynthesis constraint would be violated.  The equalisation of temperatures between our and mirror worlds occurs during one Hubble time near $T\sim M$ (even if it has not occurred early). Therefore, the physics of the  multidimensional universe at $T\ll M$ is not very important. We can use ordinary 4D physics near $T\sim M$ for estimates.

The effect of the temperatures equalization between our and mirror worlds should operate in the case when the worlds are almost symmetrical, but have a small temperature difference \cite{BerDolMoh96}. In these models, there are strong constraints on the number of relativistic degrees of freedom based on primordial nucleosynthesis. For strongly  asymmetric worlds the equalization effect can also work, but for quantitative calculations the different approaches are required, which are beyond the scope of our work. 

Let us conclude with a few comments.

1. Entropy transfer. As it was noted in \cite{Foo14}, although there is a balance of energy, the total entropy increases, because $\delta Q_2=-\delta Q_1$, but at the different temperatures $\delta Q_1/T_1\neq-\delta Q_2/T_2$. The increase of entropy occurs in the same way as the entropy increases when the temperatures of two bodies initially having different temperatures are equalized. In \cite{CarGla87,CiaFoo08,FooVag15} the mixing of our photons and mirror world photon was considered. With this mixing, the entropy in the intermediate states is not delayed. Our variant with BHs is more interesting in this respect, since it is known, that BHs themselves carry entropy, and the BH entropy is expressed through its horizon area by known formulas. Therefore, it is interesting to consider the two questions: how much entropy a BH transfers between worlds in comparison with own BH's entropy and how much entropy is enclosed in BHs at every cosmological instant of time. The last question has sense because the BHs evaporate not instantaneously, but have a certain lifetimes. One should take into account that black holes can born with relativistic velocities, therefore their energy $Mc^2/\sqrt{1-v^2/c^2}$ can exceed the rest energy $Mc^2$. However, the BH motion does not affect its entropy as in the case of the moving medium \cite{LanLif80}.

2. Planckeons. The remnants of primordial BHs were considered in many works in different aspects. In particular, the remnants can help solv the information loss paradox \cite{CheOngYeo15}. The remnants of the micro BHs can form at the particles collisions (not primordial) in the early universe  \cite{BarFerGra05,ConWiz07}, \cite{NakYok18}, \cite{Baretal19}. In the case the black holes leave stable remnants (Planckeon) the fate of the multi-dimensional universe would be dramatic not only in the mirror matter models but even for single particle sector because the universe goes into the dust-like stage very early. 

3. Primordial BHs. The evaporation of primordial BHs can also be considered as a canal between our and mirror worlds (especially the region of their masses $<10^9$~g). Equalization of the temperatures in this case provides new constraints on the primordial black holes at a small mass region. One can assume that in the early epoch the primordial BHs begin to dominate in density, then evaporate, and all was thermalized. In ordinary cosmology, this would have consequences for entropy generation \cite{ZelSta76}. In models with mirror matter, due to the evaporation of primordial BHs, the temperature asymmetry between our and the mirror world will be destroyed. Thus, it is possible to obtain new constraints on the primordial BHs in models with mirror matter in comparison with the known entropy bounds on primordial BHs \cite{ZelSta76}. Microscopic primordial BHs may arise from the preheating instability and subsequently dominate the content of the Universe, and their evaporation may be the source of reheating \cite{MarPapVen19,Maretal20,AucVen20}.

\section*{Acknowledgements}
The authors are grateful to Z.~Berezhiani and A.~Gazizov for useful discussions and to the anonymous Referee for valuable comments. The research by M.K.was financially supported by Southern Federal University, 2020 Project VnGr/2020-03-IF.

\end{document}